\documentclass[runningheads]{llncs}
\usepackage[T1]{fontenc}
\usepackage{multirow}
\usepackage{graphicx}
\usepackage{subcaption} 
\usepackage{float}
\usepackage{tcolorbox}
\tcbuselibrary{listings} 
\usepackage{booktabs}
\usepackage{tabularx}

\usepackage[dvipsnames]{xcolor}
\definecolor{ForestGreen}{rgb}{0.13, 0.55, 0.13}

\usepackage{soul,color}
\soulregister\cite7
\soulregister\ref7
\soulregister\pageref7
\setstcolor{red}
\setul{}{.2ex}

\usepackage{import} 
\usepackage{amsmath}
\usepackage{hyperref}

\hypersetup{
    colorlinks=true,
    linkcolor=blue,
    citecolor=blue,
    urlcolor=blue
}
\usepackage[section]{placeins}

\usepackage{enumitem}
\usepackage[numbers, sort]{natbib}

\begin{document}
%
\title{
The Effect of Document Summarization on LLM-Based Relevance Judgments%
}
\titlerunning{Document Summarization for LLM-Based Relevance Judgments}
%
\vspace{-25mm}
\author{Samaneh Mohtadi\inst{1} \and
Kevin Roitero\inst{2} \and 
Stefano Mizzaro\inst{2} \and Gianluca Demartini\inst{1}}
\authorrunning{S. Mohtadi et al.}
%
\vspace{-25mm}
\institute{The University of Queensland, Brisbane QLD 4072, Australia 
\email{s.mohtadi@uq.edu.au, g.demartini@uq.edu.au}\\
\and
University of Udine, Via delle Scienze, 206, 33100 Udine, Italy
\email{kevin.roitero@uniud.it, mizzaro@uniud.it}\\}
\maketitle              
\vspace{-5mm}
\begin{abstract}
Relevance judgments are central to the evaluation of Information Retrieval (IR) systems, but obtaining them from human annotators is costly and time-consuming. Large Language Models (LLMs) have recently been proposed as automated assessors, showing promising alignment with human annotations. 
Most prior studies have treated documents as fixed units, feeding their full content directly to LLM assessors. 
We investigate how text summarization affects the reliability of LLM-based judgments and their downstream impact on IR evaluation.
Using state-of-the-art LLMs across multiple TREC collections, we compare judgments made from full documents with those based on LLM-generated summaries of different lengths. We examine their agreement with human labels, their effect on retrieval effectiveness evaluation, and their influence on IR systems' ranking stability.
Our findings show that summary-based judgments achieve comparable stability in systems' ranking to full-document judgments, while introducing systematic shifts in label distributions and biases that vary by model and dataset. These results highlight summarization as both an opportunity for more efficient large-scale IR evaluation and a methodological choice with important implications for the reliability of automatic judgments.

\end{abstract}
\section{Introduction}
Evaluation in Information Retrieval (IR) fundamentally relies on relevance judgments, the decisions made by human assessors regarding to what extent a document satisfies an information need. These judgments form the basis for evaluating and ranking search systems~\cite{lawrence1998pagerank,pan2007google}. However, collecting them at scale is both costly and time-consuming. For instance, a typical TREC track demands a team of six trained assessors working full-time for several weeks~\cite{soboroff2025don}. 
Recent advances in Large Language Models (LLMs) offer a potential alternative to manual judging by generating relevance labels at scale~\cite{faggioli2023perspectives,faggioli2024determines,li2025generation,zhuang2024beyond}. Compared to human assessors, including crowd workers, LLMs reduce costs and turnaround time while hopefully maintaining sufficient alignment with human labels to preserve system comparisons~\cite{thomas2024large,balog2025rankers}. Empirical studies further suggest that LLM-based judgments are less affected by context switching and often produce more consistent outcomes~\cite{frobe2025large}. 

At the same time, the form of evidence presented to the assessor plays a crucial role. Full-document judgments provide rich context but demand significant resources, making summarization an attractive alternative. Summarization is one of the core abilities of modern LLMs~\cite{brown2020language,openai2023gpt,min2025towards,rehman2025evaluating}, and initial studies suggest that summary-based evidence can provide little loss in accuracy compared to full documents~\cite{roitero2025efficiency}. However, summaries also represent a form of semantic compression that comes with a risk of bias or distorted evaluation outcomes~\cite{zhang2024benchmarking,peters2025generalization,wan2024positional,askari2024assessing}.
In IR, prior work on LLM assessors has concentrated mainly on full-document judgments~\cite{balog2025rankers,faggioli2023perspectives,thomas2024large}, and existing summarization evaluation frameworks do not address how summarized evidence affects LLM-based relevance judgments. 

In this paper, we present the first systematic study comparing summary-based and full-document LLM relevance judgments across three popular TREC datasets. We analyze how judgments under these two modalities differ in their agreement with human labels, their impact on system effectiveness evaluation, and on systems' ranking stability (i.e., how systems are ranked on the basis of their effectiveness). We also examine the reliability of different summary lengths and associated costs. Specifically, we address the following research questions:
\begin{enumerate}[label={RQ\arabic*},leftmargin=*]
    \item \label{rq:agreement} How do relevance judgments from LLMs and humans compare in label distribution and agreement  under full-document vs summary modalities?
    
    \item \label{rq:ranking} How do summary-based relevance judgments affect system effectiveness evaluation and the stability of effectiveness-based systems ranking?
    
    \item  \label{rq:compression} How do different levels of semantic compression (i.e., summary lengths) affect the reliability and consistency of LLM-based relevance judgments?
\end{enumerate}

\section{Background and Related Work}

Test-collection–based evaluation has underpinned IR since the Cranfield experiments, where documents, topics, and human relevance judgments (qrels) enable reproducible system comparison~\cite{cleverdon1967cranfield}. While early studies showed that assessor disagreement is common, system rankings are generally robust to such variability~\cite{voorhees1998variations,voorhees2000variations,zobel1998reliable}. Later work examined assessor effects, pooling bias, and the robustness of evaluation metrics under incomplete judgments, reinforcing both the strengths and limitations of human-based evaluation~\cite{Carterette2010AssessorError,Buckley2004RetrievalIncomplete,Sakai2008IRMetricsIncomplete,webber2010effect}.
These studies established the foundations of IR evaluation practice, with human relevance judgments serving as the reference standard against which automated approaches must be validated.

Recent research has investigated LLMs as automated relevance assessors, with the potential to supplement or even replace human judges. Surveys of the emerging LLM-as-a-judge paradigm provide systematic analyses of its functionality, methodology, applications, and limitations~\cite{li2024llms,faggioli2023perspectives}. Empirical IR studies have shown that LLM judgments often align closely with human labels and largely preserve system comparisons, both under full-document evidence and in TREC-style or commercial retrieval settings~\cite{balog2025rankers,thomas2024large,upadhyay2025large,faggioli2023perspectives,faggioli2024determines}. Large-scale experiments show that LLM assessors scale efficiently and are often more self-consistent than human annotators~\cite{frobe2025large,arabzadeh2025benchmarking}. The LLMJudge challenge (SIGIR 2024) extended this line of work by releasing LLM-based relabelings of TREC Deep Learning judgments to study assessor bias, prompt sensitivity, and model selection effects~\cite{rahmani2025judging}. At the same time, several works caution against over-reliance on LLM assessors. Clarke and Dietz~\cite{clarke2024llm} and Dietz et al.~\cite{dietz2025llm} argue that LLM judgments risk bias and lack transparency, while Soboroff~\cite{soboroff2024dont} emphasizes the need for careful methodology to avoid misleading outcomes. These critiques suggest that LLM reliability depends heavily on prompt design, dataset context, and evaluation protocols, yet nearly all prior work assumes full-document evidence, leaving open whether judgments remain robust under summarization.
 
In parallel, summarization research has developed evaluation frameworks that emphasize dimensions such as faithfulness, completeness, and conciseness~\cite{song2024finesure}. Benchmarks such as SummEval and G-Eval reveal the shortcomings of lexical metrics and the advantages of LLM-based evaluation~\cite{fabbri2021summeval,liu2023g}, while FineSurE and MSumBench extend to fine-grained and cross-domain settings~\cite{song2024finesure,min2025towards}. Despite this progress, challenges remain since evaluation outcomes are highly sensitive to reference quality~\cite{gigant2024mitigating}, and models often generalize poorly across domains~\cite{liu2023benchmarking,peters2025generalization}. Other research shows that, in a crowdsourcing setting, summaries can improve assessor efficiency at the cost of little accuracy reduction~\cite{roitero2025efficiency}. Similar studies in news and political summarization show that evaluation outcomes are   sensitive to reference style and can reflect systematic framing biases~\cite{fabbri2021summeval,zhang2024benchmarking,onishi2024political,bang2024measuring}. These findings suggest that while summarization may offer efficiency gains, it also risks distorting evaluation outcomes when applied in IR settings. 
Overall, prior work shows that LLMs are reliable assessors with evidence under full document conditions alongside notable concerns about bias and transparency. Summarization research highlights efficiency gains but also risks of information loss and distortion, leaving open how semantic compression affects LLM-based relevance judgments and their downstream impact on IR evaluation.

\section{Methodology and Experimental Setting}
We outline our methodology and setting for comparing summary-based and full-document LLM judgments. For full reproducibility, we will release all code and relevant material upon acceptance.

\begin{figure}[tb]
\centering
\begin{tcolorbox}[colback=white,colframe=black, boxrule=.5pt,
boxsep=1pt,left=1pt,right=1pt,top=1pt,bottom=1pt, 
width=1\linewidth]%
\scriptsize{
You are a professional summarizer. Your task is to create a concise and accurate summary of the following document.\\
Constraints:\\
- Only produce the summary Text\\
- Do not offer extra assistance\\
- Avoid adding headings or topics\\
- Summarize the document content only avoid paraphrasing.\\
- Do NOT use any information outside the document.\\
- No opinions; just the document’s information.\\
- Preserve key facts, entities, dates, numbers, and names exactly as written.\\
- Focus on the main ideas and essential information; omit boilerplate or formatting artifacts.\\
- If the document is empty, respond with: $\text{NO\_CONTENT}$.\\
- Include who/what/when/where/quantities if present.\\
- The number of words in summary must be smaller than the number of words of the document\\
- Minimize redundancy by referring to repeated terms only once unless essential for clarity\\
- Give me a summary at maximum about <$N$> tokens.\\
Document: <DOC>
}
\end{tcolorbox}\vspace{-1em}
\caption{Prompt for generating summaries.}
\label{fig:summarization-prompt}
\end{figure}
\subsection{Datasets}
We evaluate on three benchmark datasets: TREC Deep Learning 2019 (DL-19)~\cite{craswell2020overviewtrec2019deep}, TREC Deep Learning 2020 (DL-20)~\cite{craswell2021overviewtrec2020deep}, and the TREC Retrieval-Augmented Generation track 2024 (RAG-24)~\cite{trec2024rag}. DL-19 and DL-20 are based on the MS MARCO passage collection~\cite{bajaj2016ms} with Bing queries covering diverse open-domain information needs and annotated by NIST assessors on a four-point graded scale. Both tracks also provide system runs, and we include all submitted runs in our ranking analysis (37 in DL-19 and 59 in DL-20). RAG-24, built on MS MARCO v2.1, applies the same grading scheme but targets retrieval-augmented generation in open-domain web search. For this track, we use the official human qrels, though system runs for the 2024 edition were not publicly available at the time of our experiments.

\subsection{Summary Generation and Validation}\label{sec:summary_generation}
To generate summaries, we used GPT-4o~\cite{openai2023gpt4} with a concise, task-agnostic prompt adapted from recent work on LLM summarization~\cite{zhang2024benchmarking,adams2023sparse}. 
The full prompt is provided in Figure~\ref{fig:summarization-prompt}.
Summaries were generated at two levels of semantic compression: 80 tokens (Summ-80) and 120 tokens (Summ-120), in addition to the full-document condition.

Since summary quality directly affects downstream judgments, we validated the summary generation approach using two complementary methods on the SummEval dataset: BERTScore~\cite{zhang2019bertscore} and G-Eval~\cite{liu2023g}. BERTScore is widely used in summarization evaluation~\cite{fabbri2021summeval,gao2025llm}, correlating better with human judgments than lexical metrics such as BLEU~\cite{papineni2002bleu} and ROUGE~\cite{chin2004rouge}. However, it tends to underestimate highly abstractive LLM outputs such as GPT-3~\cite{goyal2022news} and GPT-4~\cite{zhang2024benchmarking}, due to paraphrasing and diverse human references. To address this limitation, we also adopt G-Eval, which leverages GPT-4 as an evaluator across coherence, consistency, fluency, and relevance, and has shown high correlation with human ratings  on the SummEval benchmark dataset~\cite{fabbri2021summeval}.
%
Using BERTScore against the 11 human references, our summaries achieved P = 0.34, R = 0.59, and F1 = 0.41. These outcomes reflect that our summaries capture much of the important content (recall) but diverge in phrasing from the gold references (precision), resulting in modest F1. This pattern aligns with prior observations that BERTScore underestimates highly abstractive LLM outputs due to paraphrasing and stylistic variation~\cite{goyal2022news,zhang2024benchmarking}.  

To complement this, we applied G-Eval. Using GPT-4o as the judge, our summaries scored Coherence = 4.52, Consistency = 4.52, Fluency = 4.74, Relevance = 4.36. (all out of 5). These values indicate high-quality summaries and are consistent with prior reports that GPT-based summarizers achieve human-preferred quality on SummEval and related benchmarks~\cite{goyal2022news,liu2023g,zhang2024benchmarking}. This suggests that our outputs are comparable in quality to recent GPT-based summarization systems and indicates that our summarization prompt produces summaries of sufficient quality, mitigating concerns about summarization artifacts and providing a reliable evidence basis for downstream IR evaluation.

\subsection{Relevance Judgments and Modalities}\label{sec:relevance-judgments}
For human relevance judgment baselines, we use the official TREC qrels, where NIST assessors applied a four-point graded scale (0 = irrelevant, 1 = related, 2 = highly relevant, 3 = perfectly relevant). These labels are treated as authoritative ground truth. For binary settings, labels 1–3 are mapped to relevant (1) and label 0 to non-relevant (0).
For LLM-based assessors, we employ GPT-4o and Llama-3.1-8B-Instruct~\cite{dubey2024llama3}. Both models are prompted in a zero-shot setting using the UMBRELA Bing-style relevance prompt~\cite{upadhyay2024umbrela}. Since the RAG-24 dataset was released in August 2024, after GPT-4o’s reported training cutoff in October 2023, there is no risk of data leakage for this collection. To ensure comparability, the parsing and evaluation pipeline is kept identical across models, so that the only varying factor is the LLM itself. 

We examine two evidence modalities for LLM-based assessment. In the full-document modality, the entire document is provided as input to the assessor. In the summary-based modality, the document is first compressed into an LLM-generated summary (Section \ref{sec:summary_generation}), which is then used as input. For summaries, no additional human annotations are collected; instead, comparisons are always made against the official TREC qrels made looking at full documents. Our study therefore isolates the role of evidence format (full vs. summary) while holding all other components constant (prompt, parser, datasets), enabling a controlled comparison of how evidence modality affects the quality and stability of IR evaluation.

\subsection{Evaluation Framework}
To compare summary-based and full-document judgments, we adopt a three-level evaluation framework, where each level targets a distinct unit of analysis.

At the local level, agreement metrics assess how consistently individual relevance labels are preserved across modalities. In the binary setting, we report Cohen’s $\kappa$~\cite{cohen1960coefficient} as the standard chance-corrected measure for two raters and Krippendorff’s $\alpha$~\cite{krippendorff2018content} for its generality and robustness to missing data. In the graded setting, we use weighted Cohen’s $\kappa$~\cite{cohen1968weighted} and Krippendorff’s $\alpha$, which account for the severity of disagreements across multiple relevance levels.

At the system level, effectiveness metrics assess how differences in judgments translate into retrieval performance. Following the official evaluation protocols of the TREC Deep Learning tracks~\cite{craswell2020overviewtrec2019deep,craswell2021overviewtrec2020deep}, we report Normalized Discounted Cumulative Gain at rank 10 (NDCG@10)~\cite{jarvelin2002cumulated} as the primary graded relevance measure. For completeness, we also report Mean Average Precision (MAP)~\cite{voorhees1998variations}, which is widely used in binary settings and provides a complementary perspective on evaluation outcomes.

At the global level, ranking stability metrics evaluate the consistency of system comparisons across modalities. We report Kendall’s $\tau$~\cite{kendall1938new} and Spearman’s $\rho$~\cite{spearman1987proof} to assess rank correlations, and Rank-Biased Overlap (RBO)~\cite{webber2010similarity} to emphasize stability among top-ranked systems.

\section{Experiments and Results}
In the following subsections, we address our three research questions: label distribution and agreement with human judgments (\ref{rq:agreement}), system effectiveness and ranking stability 
under summary-based judgments (\ref{rq:ranking}) and the impact of semantic compression across different summary lengths (\ref{rq:compression}).

\subsection{\ref{rq:agreement}: Label Distribution and Agreement}
\subsubsection{Label Distribution Analysis.}
Table~\ref{tab:reljudgement_distributions} compares the distribution of relevance judgments across modalities. For GPT-4o, summarization length has little effect, where Summ-80 and Summ-120 produce very similar distributions. Both settings bring the share of non-relevant labels (0) closer to the official baseline compared to full documents, primarily by reducing label-1 share. This indicates that summarization tends to collapse fine distinctions between non-relevant and marginally relevant documents, but does so consistently across DL-19, DL-20, and RAG-24.  

Llama-3.1-8B, by contrast, shows greater sensitivity to compression. While it mirrors GPT-4o’s shift from label-1 to label-0 in DL-19, the effect diverges on larger datasets: in DL-20, summaries reduce label-0 while inflating labels 1 and 2, and in RAG-24, 120-token summaries overproduce label-1. These instabilities suggest that Llama-3.1-8B struggles to preserve stable label distributions under compression, particularly in more complex collections.  

In sum, summarization exerts a mild and predictable effect on GPT-4o but destabilizes Llama-3.1-8B, highlighting how input compression interacts differently with model capacity. These findings contribute to answering \ref{rq:agreement} by showing that while GPT-4o remains robust to summarization, smaller models may distort label distributions when input is semantically compressed.

\begin{table}[tb]
\centering
\scriptsize
\caption{Distribution of relevance judgments (in \%) across DL-19, DL-20, and RAG-24, comparing human annotations with LLM assessors.}
\label{tab:reljudgement_distributions}
\begin{tabular}{l l
                cccc
                cccc
                cccc}
\toprule
\textbf{Annotator} & \textbf{Modality} &
\multicolumn{4}{c}{\textbf{DL-19}} &
\multicolumn{4}{c}{\textbf{DL-20}} &
\multicolumn{4}{c}{\textbf{RAG-24}} \\
\cmidrule(lr){3-6} \cmidrule(lr){7-10} \cmidrule(lr){11-14}
 & & 0 & 1 & 2 & 3
 & 0 & 1 & 2 & 3
 & 0 & 1 & 2 & 3  \\
\midrule
GPT-4o & Summ-80
& 50.0 & 31.4 & 9.9 & 8.7
& 60.6 & 25.3 & 7.6 & 6.5
& 26.1 & 52.4 & 16.3 & 5.2 \\

       & Summ-120
& 49.7 & 31.6 & 10.0 & 8.7
& 59.8 & 26.3 & 7.4 & 6.5
& 26.2 & 51.3 & 16.9 & 5.6 \\

       & Full Document
& 46.6 & 33.2 & 11.1 & 9.1
& 56.7 & 28.2 & 8.4 & 6.7
& 21.1 & 57.1 & 18.6 & 3.2 \\
[0.75ex]
\hline \\[-0.25ex]
Llama-3.1-8B & Summ-80
& 55.4 & 1.8 & 30.7 & 12.1
& 30.0 & 12.1 & 47.7 & 10.2
& 13.3 & 23.1 & 45.0 & 18.7 \\

       & Summ-120
& 55.2 & 2.2 & 30.2 & 12.4
& 62.7 & 2.5 & 26.4 & 8.4
& 35.9 & 7.6 & 45.6 & 10.9 \\

       & Full Document
& 46.5 & 4.2 & 31.1 & 18.2
& 54.0 & 4.8 & 27.7 & 13.5
& 21.0 & 6.8 & 47.2 & 25.1 \\

\midrule
Official &   
& 55.0 & 17.3 & 19.5 & 7.5
& 68.3 & 17.0 & 9.0 & 5.7
& 37.3 & 31.1 & 23.0 & 8.6 \\

\bottomrule
\end{tabular}
\end{table}

\subsubsection{Relevance Judgment Analysis.}
Tables~\ref{tab:agreement-graded} and \ref{tab:agreement-binary} report graded and binary agreement scores between LLM judgments and official human labels. For graded judgments (Table~\ref{tab:agreement-graded}), GPT-4o achieves higher agreement than Llama-3.1-8B across all datasets and metrics. Weighted $\kappa$ values remain stable across modalities, with the strongest performance in DL-19 where both summaries and full documents exceed 0.58. Krippendorff’s $\alpha$ also shows higher alignment, particularly in DL-19 and DL-20. Importantly, GPT-4o under summarization consistently achieves higher scores than in the full-document condition, indicating that semantic compression does not harm and can even enhance graded agreement. By contrast, Llama-3.1-8B records substantially lower agreement but shows relative improvements under summarization. Its best $\kappa$ values occur with Summ-80 on DL-19, while Summ-120 yields higher scores on DL-20 and RAG-24. However, $\alpha$ values remain very low overall, in some cases approaching zero, indicating difficulty in capturing fine-grained distinctions.

\begin{table}[tb]
\centering
\scriptsize
\caption{Agreement analysis (Graded labels).}
\label{tab:agreement-graded}
\begin{tabular}{ll
                ccc  
                ccc} 
\toprule
\textbf{LLM judge} & \textbf{Modality} &
\multicolumn{3}{c}{\textbf{Weighted Cohen's $\kappa$}} &
\multicolumn{3}{c}{\textbf{Krippendorff's $\alpha$}} \\
\cmidrule(lr){3-5} \cmidrule(lr){6-8}
& & \textbf{DL-19} & \textbf{DL-20} & \textbf{RAG-24}
  & \textbf{DL-19} & \textbf{DL-20} & \textbf{RAG-24}
   \\
\midrule
GPT-4o & Summ-80        & \textbf{.583} & .561 & .493 & .331 & .348 & .235 \\
       & Summ-120       & .574 & \textbf{.562} & \textbf{.497} & .326 & \textbf{.350} & \textbf{.240} \\
       & FullDocument   & .582 & .560 & .493 & \textbf{.340} & .341 & .225 \\
[0.75ex]
\hline \\[-0.25ex]
Llama-3.1-8B & Summ-80    & \textbf{.362} & .245 & .165 & \textbf{.193} & .003 & .004 \\
           & Summ-120   & .350 & \textbf{.313} & \textbf{.247} & .188 & \textbf{.141} & \textbf{.057} \\
           & FullDocument & .335 & .277 & .198 & .155 & .114 & .010 \\
\bottomrule
\end{tabular}
\end{table}
For binary judgments (Table~\ref{tab:agreement-binary}), obtained using the binarization procedure outlined in Section~\ref{sec:relevance-judgments}, GPT-4o again consistently outperforms LLaMA-3.1-8B across both $\kappa$ and $\alpha$. Its best scores come from summaries, with Summ-80 performing slightly better than Summ-120 across all datasets and metrics. This confirms that even shorter summaries do not reduce agreement with human annotations and can slightly improve it. In contrast, LLaMA-3.1-8B  shows consistently lower agreement scores than GPT-4o. It achieves relatively higher values with Summ-80 on DL-19 for both $\kappa$ and $\alpha$, but this improvement is limited compared to GPT-4o. On DL-20 and RAG-24, the pattern differs, with $\kappa$ peaking with full documents, while $\alpha$ peaks under Summ-120. Despite these fluctuations, Llama-3.1-8B remains well below GPT-4o in every condition. 

\begin{table}[tb]
\centering
\scriptsize
\caption{Agreement analysis (Binary labels)}
\label{tab:agreement-binary}
\begin{tabular}{ll
                ccc  
                ccc}  

\toprule
\textbf{LLM judge} & \textbf{Modality} &
\multicolumn{3}{c}{\textbf{Cohen's $\kappa$}} &
\multicolumn{3}{c}{\textbf{Krippendorff's $\alpha$}} \\
\cmidrule(lr){3-5} \cmidrule(lr){6-8} 
& & \textbf{DL-19} & \textbf{DL-20} & \textbf{RAG-24}
  & \textbf{DL-19} & \textbf{DL-20} & \textbf{RAG-24} \\
\midrule
GPT-4o & Summ-80           & \textbf{.524} & \textbf{.490} & \textbf{.397} & \textbf{.522} & \textbf{.487} & \textbf{.389}  \\
       & Summ-120          & .518 & .488 & .394 & .516 & .484 & .386  \\
       & FullDocument      & .517 & .481 & .396 & .513 & .474 & .377  \\ 
\hline 
Llama-3.1-8B & Summ-80       & \textbf{.320} & .215 & .101 & \textbf{.320} & .100 & .032  \\
           & Summ-120      & .311 & .291 & .181 & .311 & \textbf{.289} & \textbf{.181}  \\
           & FullDocument  & .301 & \textbf{.326} & \textbf{.381} & .295 & .255 & .120  \\
\bottomrule
\end{tabular}
\end{table}
Overall, these results address \ref{rq:agreement} by showing that GPT-4o maintains closer alignment with human judgments and stable performance under summarization, while LLaMA-3.1-8B is less consistent, particularly for graded relevance.

\subsection{\ref{rq:ranking}: System Effectiveness and Ranking Stability}

\subsubsection{System Effectiveness Analysis.}
\begin{figure}[tb]
  \centering
  \begin{tabular}{cc}
    \includegraphics[width=.38\linewidth]{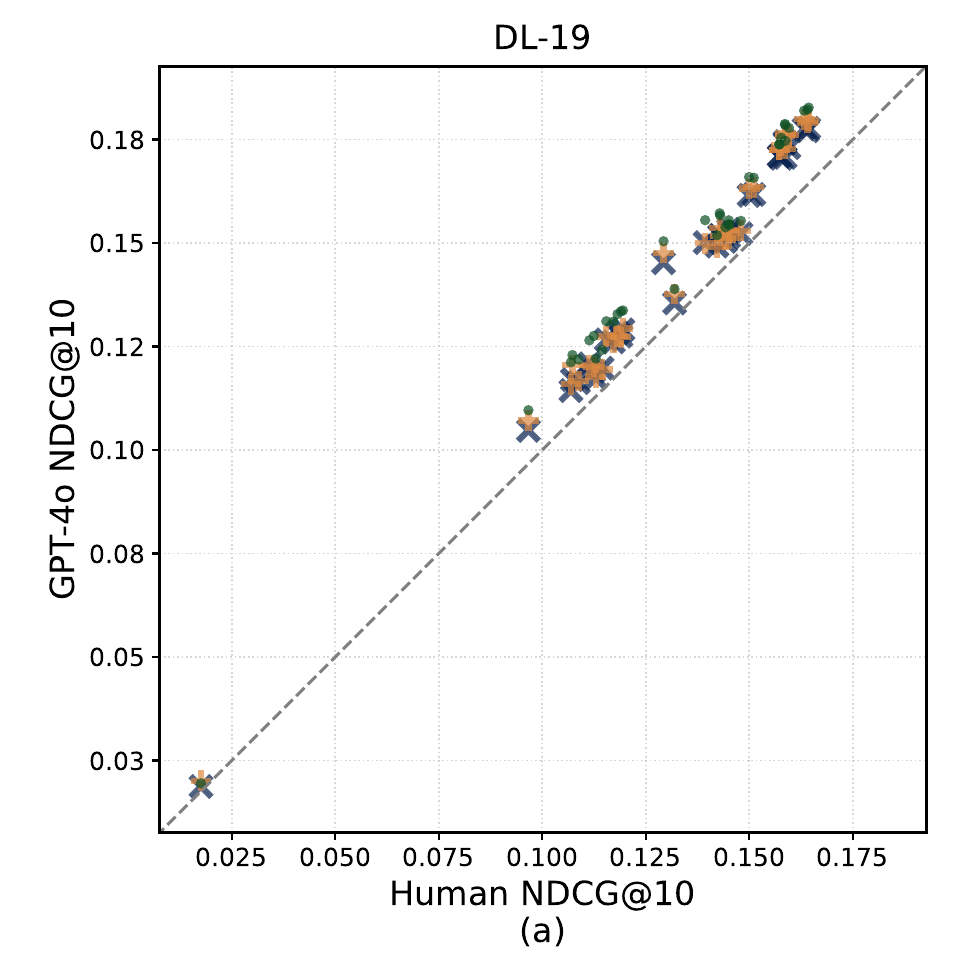} &
    \includegraphics[width=.38\linewidth]{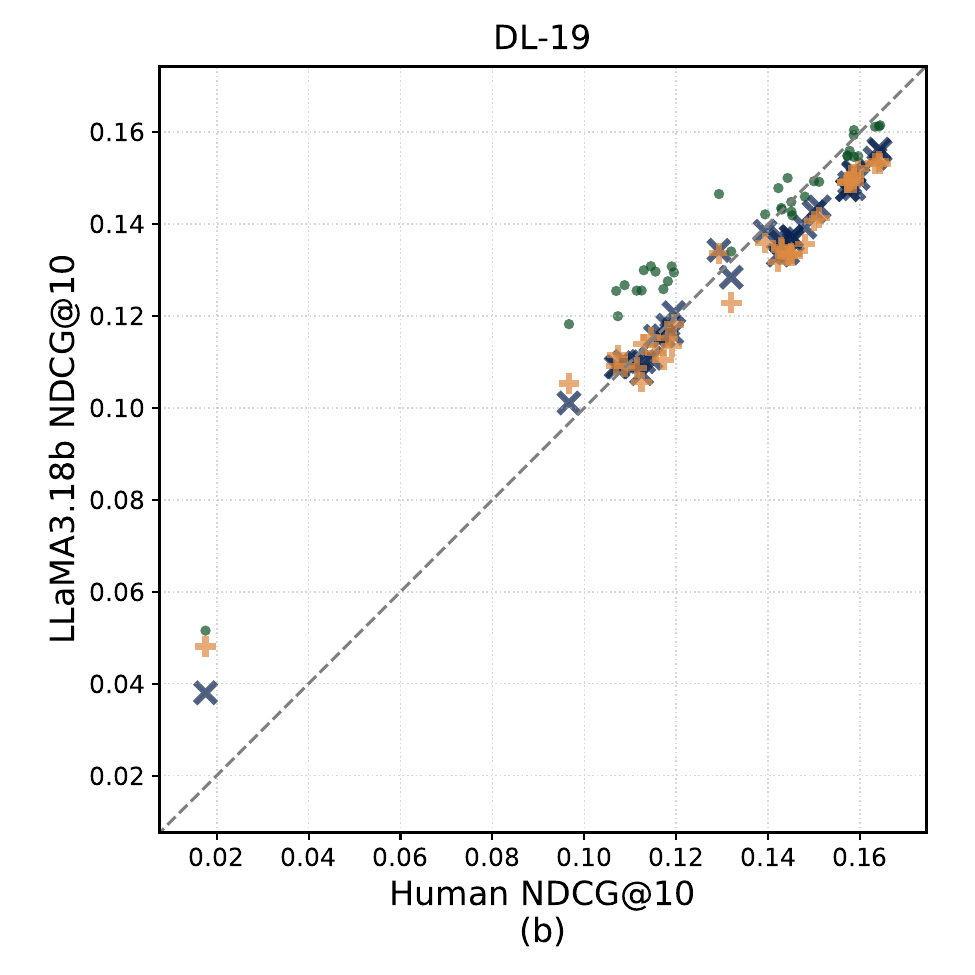}\\
    \includegraphics[width=.38\linewidth]{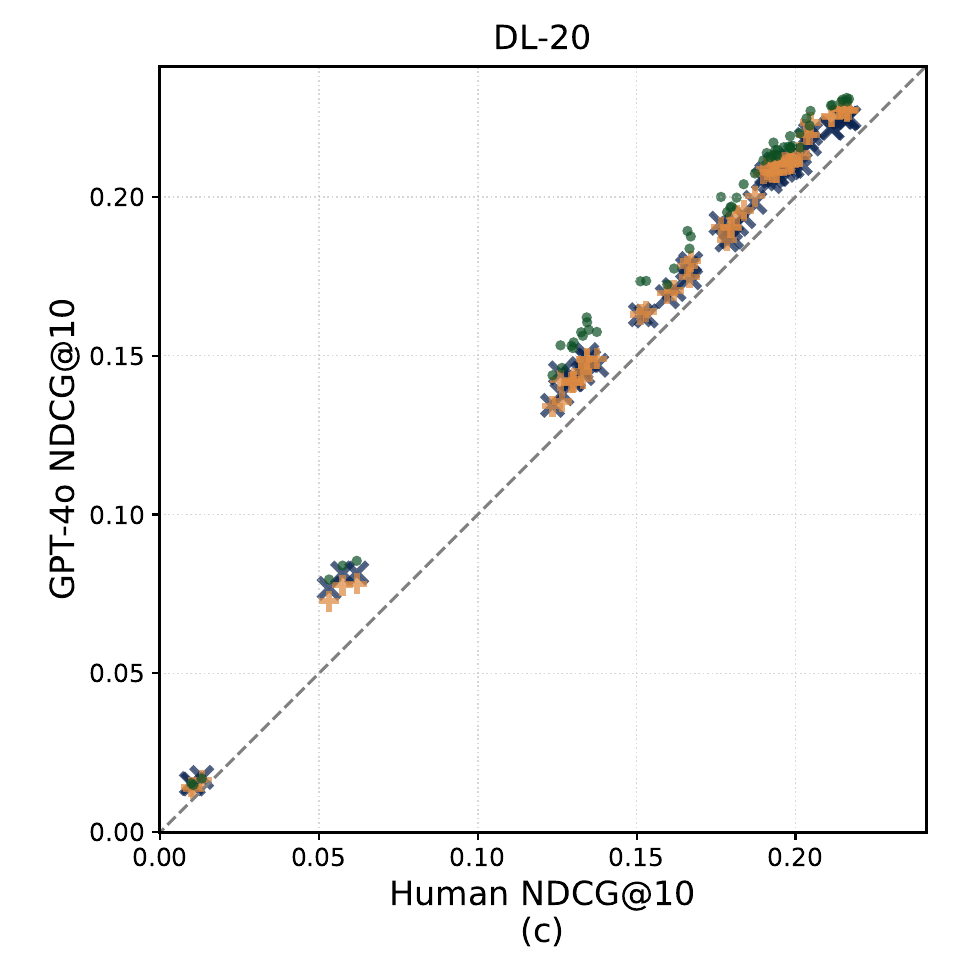} &
    \includegraphics[width=.38\linewidth]{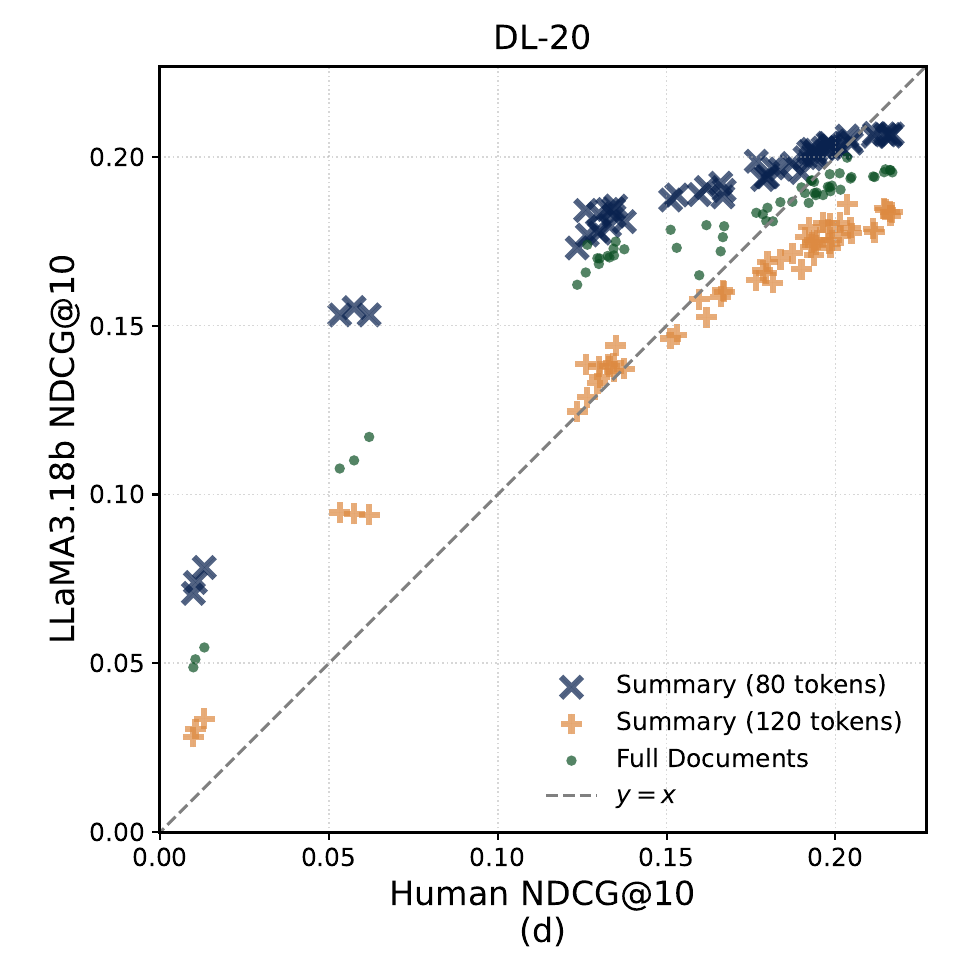}\\[-1em]
  \end{tabular}
  \caption{Scatter plots of Human- vs.\ LLM-derived retrieval effectiveness (NDCG@10) for DL-19 and DL-20. 
}
  \label{fig:ndcg-scatter}
\end{figure}

\begin{figure}[tb]
  \centering
  \begin{tabular}{cc}
    \includegraphics[width=.38\linewidth]{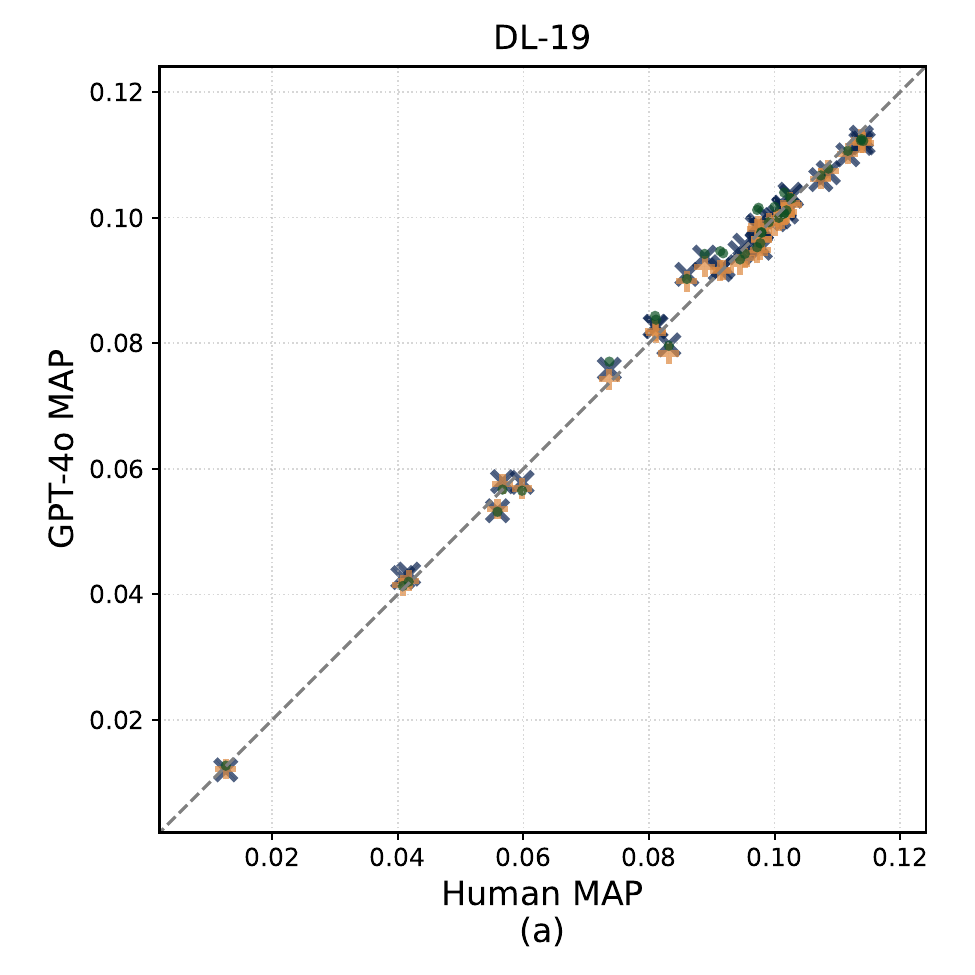} &
    \includegraphics[width=.38\linewidth]{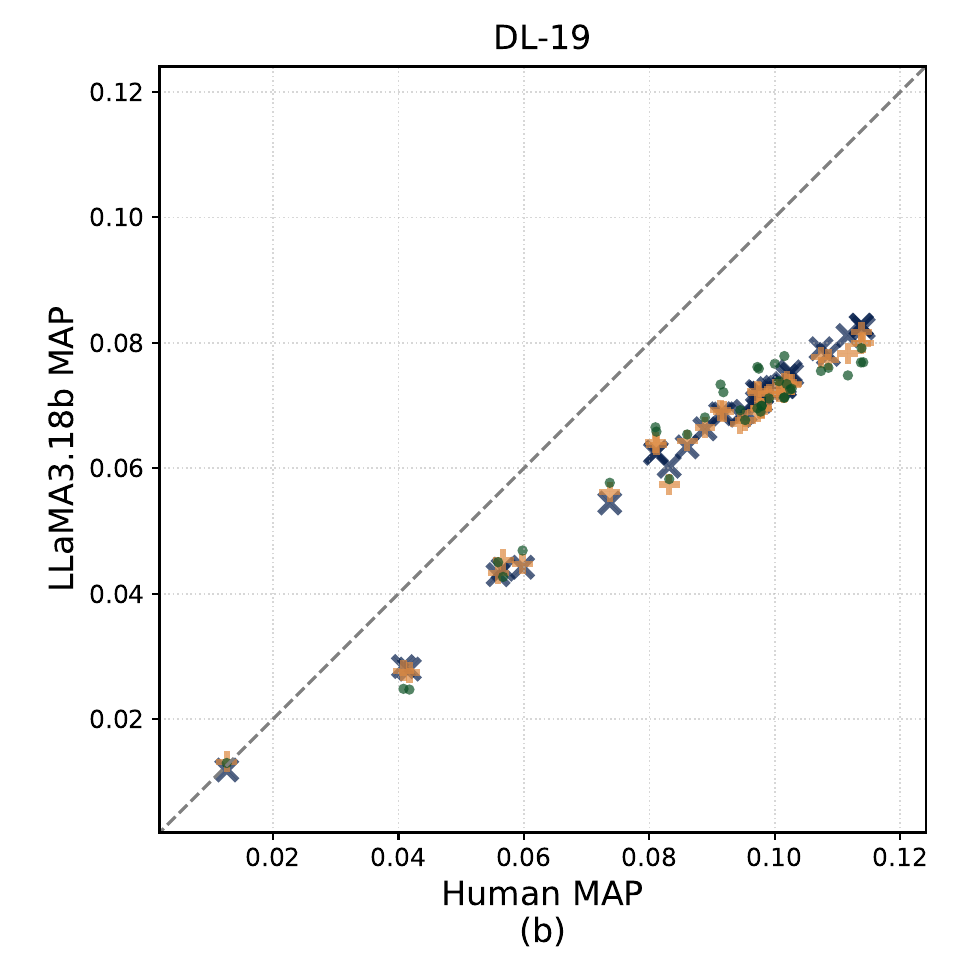}\\
    \includegraphics[width=.38\linewidth]{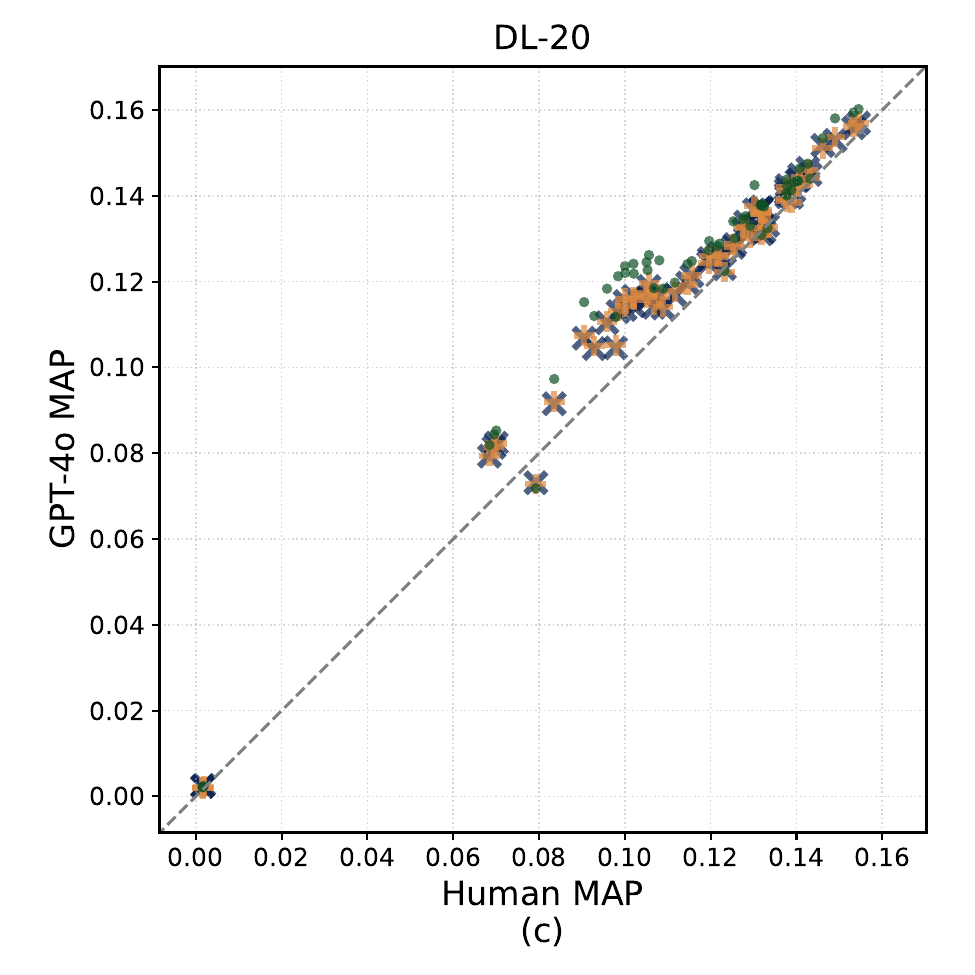} &
    \includegraphics[width=.38\linewidth]{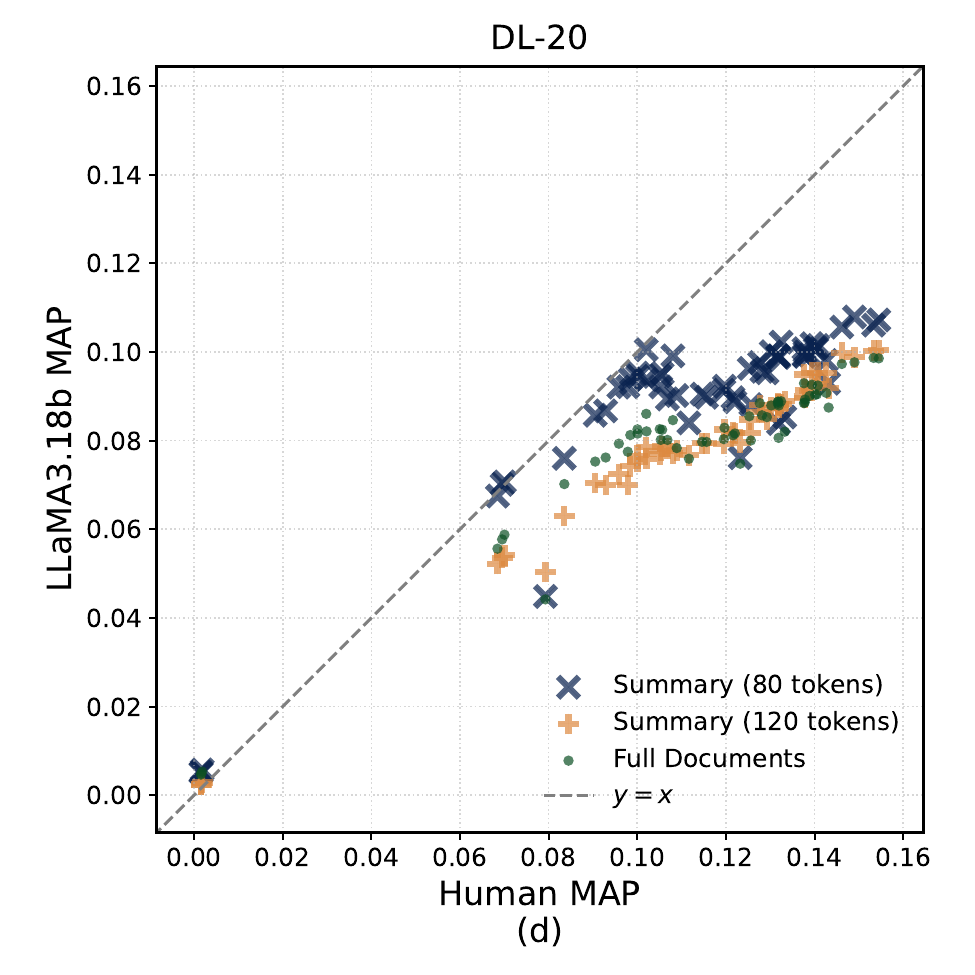}\\[-1em]
  \end{tabular}
   \caption{Scatter plots of Human- vs.\ LLM-derived retrieval effectiveness (MAP).}
  \label{fig:map-scatter}
\end{figure}

To assess how summary and full-document judgments affect retrieval evaluation, we compare system effectiveness scores against official human labels. Figures~\ref{fig:ndcg-scatter} and \ref{fig:map-scatter} plot NDCG@10 and MAP for DL-19 and DL-20, where each point corresponds to a submitted run evaluated with human judgments (x-axis) and LLM-derived judgments (y-axis). Separate markers denote summaries of 80 and 120 tokens, and full-document conditions, with diagonal $y=x$ line indicating perfect preservation relative to human labels. 

For NDCG@10, GPT-4o maintains close agreement with human scores across both datasets (Figure~\ref{fig:ndcg-scatter}(a) and \ref{fig:ndcg-scatter}(c)). Interestingly, summaries (80/120 tokens) often track human labels more tightly than full-document, though they sit slightly above the diagonal, reflecting mild score inflation. This suggests that summarization reduces distracting context and sharpens relevance signals, which improves alignment with human judgments, while slightly overestimating effectiveness.
LLaMA-3.1-8B shows more variable behavior. In DL-19 (Figure~\ref{fig:ndcg-scatter}(b)), summaries bring some runs closer to human labels, while in DL-20 (Figure~\ref{fig:ndcg-scatter}(d)), summary- and full-document scores scatter widely, indicating that summarization can both mitigate and amplify instability depending on dataset complexity.

MAP highlights how ranking sensitivity varies by judging modality. GPT-4o again produces scores closely aligned with human labels across datasets (Figures \ref{fig:map-scatter}(a) and \ref{fig:map-scatter}(c)) with deviations concentrated among weaker runs. LLaMA-3.1-8B (Figure~\ref{fig:map-scatter}(b) and \ref{fig:map-scatter}(d)) diverges more strongly, especially in DL-20, where nearly all runs fall below the diagonal. Interestingly, 80-token summaries align more closely with human labels than either 120-token summaries or full-document judgments, suggesting that shorter summaries capture the most relevant evidence more effectively by filtering noise. Dataset characteristics likely explain this. DL-19 contains shorter passages, so it preserves key evidence in summaries, while DL-20's longer documents make summarization prone to dropping some context, leading to systematic underestimation. 

Overall, GPT-4o preserves system effectiveness scores across modalities with only minor inflation, while LLaMA-3.1-8B is more sensitive to dataset complexity and summarization length. 
\subsubsection{Ranking Stability Analysis.}
\begin{figure}[tbp]
  \centering
   \includegraphics[width=\linewidth]{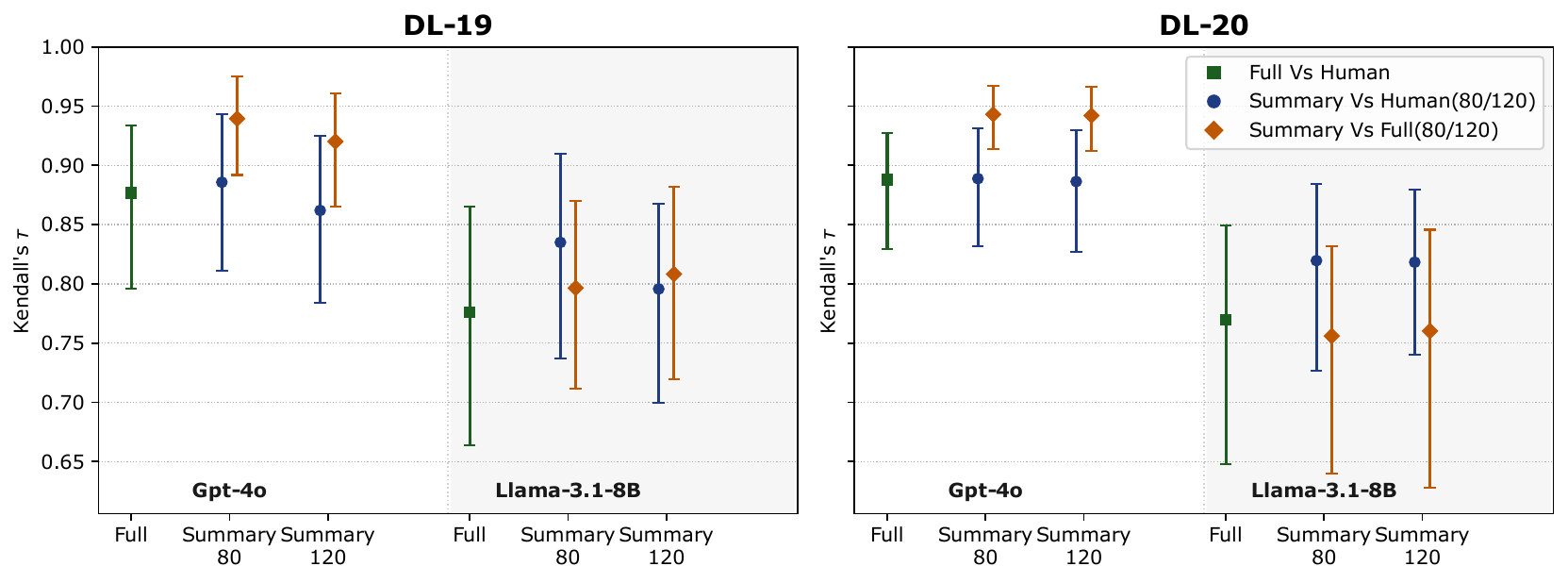}
    \caption{Kendall’s $\tau$ (computed on nDCG@10 system scores) with 95\% bootstrap confidence intervals across judging modalities.}

  \label{fig:kendall_ci}
\end{figure}

We analyze the stability of systems rankings under different judging modalities to assess whether summary-based relevance labels preserve reliable system comparisons. Figure~\ref{fig:kendall_ci} provides a high-level view, showing Kendall’s $\tau$ correlations on NDCG@10 rankings with 95\% confidence intervals of 2000 times bootstrapping. This captures both average agreement and variability across inputs. Table~\ref{tab:system_stability} complements this view by reporting Kendall’s $\tau$, Pearson’s $\rho$, and RBO across datasets and metrics.

As shown in Figure~\ref{fig:kendall_ci}, GPT-4o achieves high stability on both DL-19 and DL-20, with Kendall’s $\tau$ values consistently above 0.9 and narrow confidence intervals across all modalities. Both Summ-80 and Summ-120 track human-based rankings as closely as full documents, confirming that summarization preserves system ordering. In contrast, LLaMA-3.1-8B is less stable, with DL-19 showing moderate agreement and wider intervals, and DL-20 exhibiting both lower central values and substantially broader intervals, indicating vulnerability to dataset complexity and summarized modalities. Summarization sometimes improves alignment with human orderings, but for LLaMA-3.1-8B it more often introduces additional uncertainty, whereas GPT-4o remains consistently stable.

The detailed results in Table~\ref{tab:system_stability}, confirm that rankings remain broadly stable, with $\tau$ and $\rho$ exceeding 0.85 in nearly all conditions. RBO values are lower, reflecting greater sensitivity at the top of the ranking. For GPT-4o, Summary-based judgments (80/120 tokens) achieve $\tau$ above 0.92 and $\rho$ above 0.96, closely matching full-document performance and aligning with human-based orderings for both MAP and NDCG@10. Summ-80 often yields the most stable configuration, particularly for MAP where it also maximizes RBO (e.g., 0.950 in DL-20). For NDCG@10, full-document judgments achieve the strongest RBO, though summaries remain competitive. These results suggest that concise summaries can deliver more reliable evaluation signals than full documents.

LLaMA-3.1-8B shows competitive stability in DL-19, where Summ-80 reaches a $\tau$ above 0.90, but performance drops sharply in DL-20 (e.g., MAP $\tau$ $\approx$ 0.738 under full documents). Summaries partially recover stability, with Summ-80 producing the best Kendall’s $\tau$ for NDCG@10, while Summ-120 yields the highest Pearson’s $\rho$. RBO trends are mixed, as summaries outperform full documents for MAP, whereas for NDCG@10 the highest overlap is obtained with full documents in DL-19 and with Summ-80 in DL-20. This indicates that summarization can mitigate instability, but its effect depends on the metric and dataset.

Overall, GPT-4o consistently delivers stable systems ranking across metrics and modalities, while LLaMA-3.1-8B remains more vulnerable to variation. Although system orderings are broadly preserved, the lower RBO values underline that the top ranks shift more depending on input modality. Nevertheless, GPT-4o preserves the highest stability at the top ranks, with Summ-80 emerging as the most reliable configuration. These findings address \ref{rq:ranking} by showing that summary-based judgments, compared to full-document ones, preserve systems ranking stability while occasionally improving robustness.
\begin{table}[tb]
\centering
\scriptsize 
\caption{System stability computed with MAP (top) and NDCG@10 (bottom).}
\label{tab:system_stability}
\begin{tabular}{ll
                cc  
                cc  
                cc} 
\toprule
\textbf{LLM judge} & \textbf{Modality} &
\multicolumn{2}{c}{\textbf{Kendall's $\tau$}} &
\multicolumn{2}{c}{\textbf{Pearson's $\rho$}} &
\multicolumn{2}{c}{\textbf{RBO}} \\
\cmidrule(lr){3-4} \cmidrule(lr){5-6} \cmidrule(lr){7-8}
& & \textbf{DL-19} & \textbf{DL-20} 
  & \textbf{DL-19} & \textbf{DL-20}
  & \textbf{DL-19} & \textbf{DL-20}  \\
\midrule
\multicolumn{8}{c}{\textbf{MAP}} \\
\midrule
GPT-4o & Summ-80           & .934 & \textbf{.924} & \textbf{.997} & \textbf{.990} & \textbf{.850} & \textbf{.950} \\
       & Summ-120          & \textbf{.946} & .915 & \textbf{.997} & .988 & .814 & .946 \\
       & FullDocument      & .886 & .874 & .995 & .975 & .771 & .927 \\ [0.75ex]
\hline \\[-0.25ex]
Llama-3.1-8B & Summ-80       & \textbf{.940} & .599 & \textbf{.997} & .904 & .801 & .695 \\
             & Summ-120      & .889 & \textbf{.917} & .991 & \textbf{.986} & \textbf{.812} & \textbf{.897} \\
             & FullDocument  & .742 & .738 & .970 & .946 & .589 & .775 \\
\midrule
\multicolumn{8}{c}{\textbf{NDCG@10}} \\
\midrule
GPT-4o & Summ-80           & \textbf{.922} & \textbf{.937} & \textbf{.996} & \textbf{.997} & .845 & .698 \\
       & Summ-120          & .910 & .929 & .994 & \textbf{.998} & .771 & .694 \\
       & FullDocument      & .892 & \textbf{.938} & .993 & .996 & \textbf{.887} & \textbf{.792} \\ [0.75ex]
\hline \\[-0.25ex]
Llama-3.1-8B & Summ-80       & \textbf{.910} & \textbf{.902} & .940 & .928 & .801 & \textbf{.648} \\
             & Summ-120      & .844 & .856 & \textbf{.986} & \textbf{.983} & .769 & .537 \\
             & FullDocument  & .829 & .850 & .984 & .946 & \textbf{.852} & .610 \\
\bottomrule
\end{tabular}
\end{table}

\subsection{\ref{rq:compression}: Impact of Semantic Compression}

Table~\ref{tab:tokens-costs} highlights the practical implications of semantic compression for scalable relevance assessment. For smaller collections such as DL-19 and DL-20 (with around 9.2K and 11.3K qrels, respectively), the cost difference between full-document and summary-based judgments is modest. Because the documents are already short, reducing them to 80–120 tokens offers limited additional savings. In contrast, RAG-24 is considerably larger, with 20,277 judged query–document pairs (and 108,479 official query–document pairs in the full release~\cite{umbrelaQrels2024}) averaging about 363 tokens each. Judging the pool of 20K pairs with full documents requires roughly 13.3M tokens, whereas using 80 and 120 token summaries reduces this to about 7.7M tokens, about 42\% reduction in the input size. Extrapolating from these statistics, if the entire RAG-24 collection were judged, the cost difference would become even more substantial. Full-document judgments would require about 39M tokens ($\approx$\$102), while summaries would cut this to around 9M ($\approx$\$23) and 13M($\approx$\$34) tokens by reduction to 80 and 120 tokens, respectively. This translates into substantial savings at scale.

This scaling behavior suggests that as collections grow and documents lengthen, the  cost of full-document judging increases rapidly. Summarization mitigates this overhead by reducing input size while preserving sufficient context for reliable judgments. Beyond financial savings, summaries also reduce the cognitive load for human assessors, who cannot feasibly process full documents at web scale. The benefits are most evident for large datasets such as RAG-24, where compression substantially lowers the annotation burden without degrading system-level outcomes. These results address \ref{rq:compression} by showing how different levels of summarization influence the scalability of LLM-based relevance judgments.

\begin{table}[tb]
\centering
\scriptsize
\caption{Comparison of token usage and cost of GPT-4o for relevance judgments.}
\label{tab:tokens-costs}
\setlength{\tabcolsep}{3pt}
\begin{tabular}{llccc ccc}
\toprule
& & \multicolumn{3}{c}{\textbf{Input Tokens (Millions)}} & \multicolumn{3}{c}{\textbf{Cost (USD)}} \\
\cmidrule(lr){3-5}\cmidrule(lr){6-8}
\textbf{Stage} & \textbf{Modality} & DL-19 & DL-20 & RAG-24 & DL-19 & DL-20 & RAG-24 \\
\midrule
\multirow{3}{*}{\shortstack{Relevance\\Judgment}}
 & Summ-80  & 3.1 & 3.8 &  7.7 &  9.0 & 11.1 & 22.5 \\
 & Summ-120 & 3.1 & 3.8 &  8.0 &  9.0 & 11.0 & 23.1 \\
 & FullDoc  & 3.2 & 4.1 & 13.3 &  9.5 & 11.7 & 35.1 \\
\bottomrule
\end{tabular}
\end{table}
\section{Conclusions and Implications}
This study examined how document summarization affects LLM-based relevance judgments across three dimensions: agreement with human labels, system effectiveness and rankings, and reliability under different levels of semantic compression. We found that GPT-4o maintains strong alignment with human judgments across modalities, often improving agreement when using summaries, while LLaMA-3.1-8B remains more sensitive to dataset complexity and compression. Summary-based judgments preserved system effectiveness scores and ranking stability, with 80-token summaries frequently emerging as the most reliable configuration. Differences between 80- and 120-token summaries were minor, suggesting that concise outputs provide sufficient context for reliable evaluation. The cost analysis further showed that semantic compression substantially reduces token and financial overhead, particularly in large-scale collections such as RAG-24, and that summaries can be reused across judges and tasks. Overall, summarization provides a scalable, cost-efficient, and faithful alternative to full-document judging, offering a practical path for future test collections and shared tasks to balance scalability, reproducibility, and fidelity.

\clearpage
\bibliographystyle{splncs04}
\bibliography{references/references} 


\end{document}